\documentclass[twocolumn,showkeys,prl,groupedaddress,amsmath,amssymb]{revtex4}
\usepackage[pdftex]{graphicx,epsfig}

\begin{document}

\title{Quantum Criticality Stabilizes High $T_c$ Superconductivity \\
Against Competing Symmetry-Breaking Instabilities}

\author{Josef Ashkenazi and Neil F. Johnson} 
\affiliation{Physics Department, University of Miami, P.O.~Box 248046, Coral Gables, FL 33126, U.S.A.}

\date{\today}

\begin{abstract}
The occurrence of high-$T_c$ superconductivity in systems including the cuprates and the iron-based superconductors, is known to coincide with the existence of anomalous normal-state properties which have been associated with quantum criticality. We argue here that this observation results from the fact that quantum criticality can allow the occurrence of very-strong-coupling superconductivity by preventing its suppression due to competing symmetry-breaking instabilities. Treating the electrons through a large-$U$ ansatz yields their {\sl separation} into boson quasiparticles which are directly involved in the formation of these instabilities, represented as their Bose condensates, and charge-carrying fermion quasiparticles which are affected by them indirectly. Within the critical regime, condensates corresponding to the different broken-symmetry states are {\sl combined}; consequently their negative effect on the pairing of the fermions is strongly diminished, enabling high-$T_c$ superconductivity to occur. The observed phase diagram of the hole-doped cuprates then derives from a hidden $T=0$ quantum phase transition between a Fermi-liquid and a non-Fermi-liquid broken-symmetry striped state. The pseudogap range within this diagram is found to include two distinct regimes, with partial pairing occurring in one of them.
\end{abstract}


\maketitle

\section{Introduction}
\begin{figure}  
\centering
\includegraphics[width=\columnwidth]{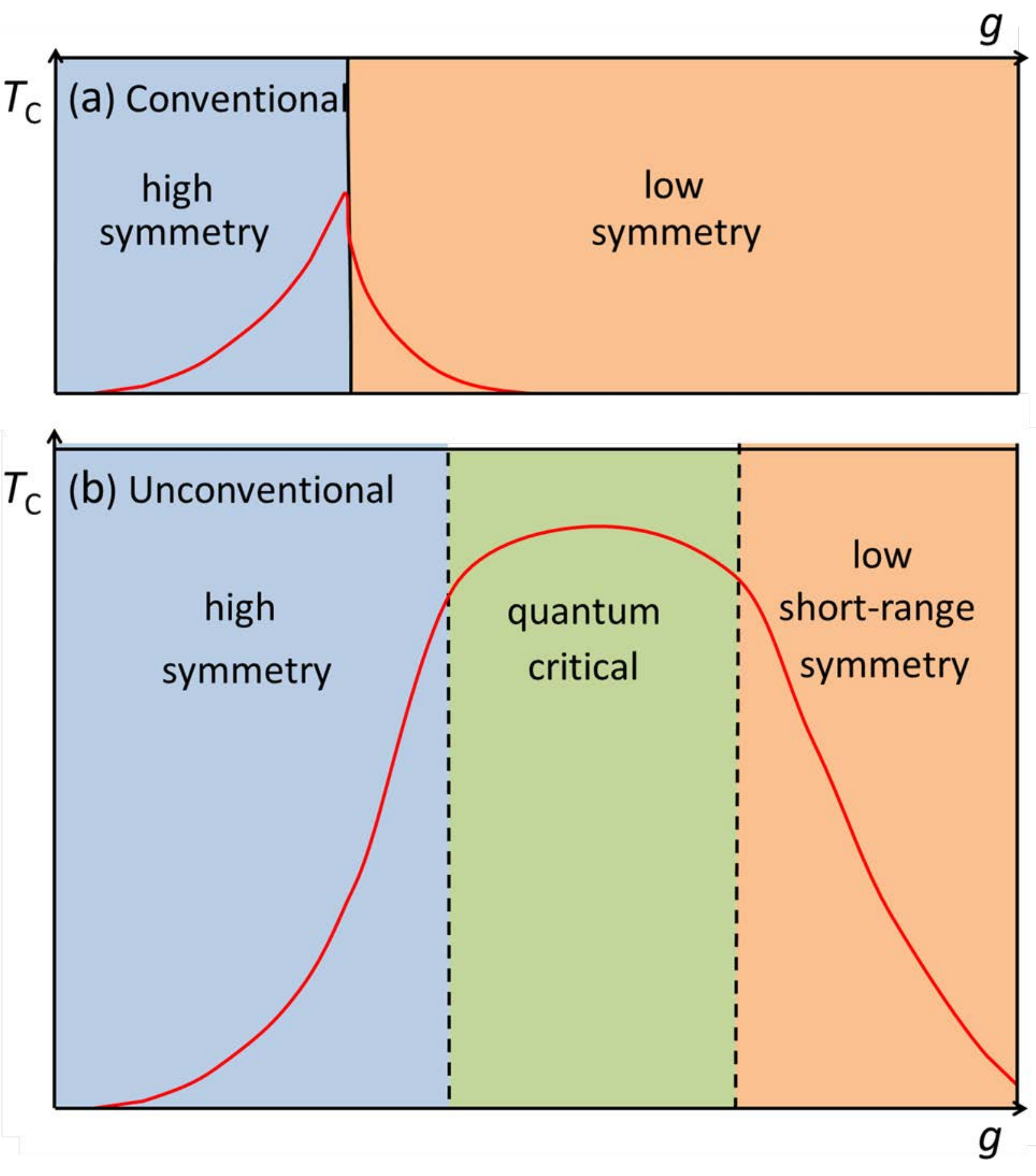}
\caption{A generic curve for the observed behavior of $T_c$ as a function of the coupling-strength parameter $g$, for (a) conventional and (b) unconventional superconductors. Solid and dashed lines represent, respectively, the boundaries of phase and gradual transitions, at $T$ just above $T_c$. This schematic figure could be further modified for specific cases due to the presence of additional phases.} 
\label{fig1}
\end{figure}

Previous attempts to raise the superconducting (SC) transition temperature $T_c$ in an effort to achieve high-$T_c$ superconductivity (HTSC),  have been hampered by the existence of competing symmetry-breaking instabilities \cite{Testardi} (e.g. of the Jahn--Teller type). These  are triggered by the coupling of the electrons to phonons or electronic excitations -- the same type of coupling which induces SC pairing -- and emerge when the coupling constants become too strong. We will use the parameter $g$ to represent the typical coupling constant strength. The symmetry breaking results in opening gaps on the Fermi surface (FS), thereby reducing the magnitude of the pairing-strength parameter $\lambda$ \cite{Kresin}.  The breakdown of time-reversal symmetry, in the case of magnetic instabilities, removes the Kramers degeneracy; this introduces a further limitation to pairing since the paired electrons stop being degenerate in the presence of disorder and impurities. Thus, a generic behavior of $T_c$ as a function of the coupling-strength parameter $g$, in what we refer to as ``conventional'' superconductors, is sketched in Fig.~\ref{fig1}(a), where the high and the low symmetry regimes  represent those of the original and the broken symmetries, mentioned above.

After the discovery of HTSC  in the cuprates, their normal state was found to be characterized by anomalous features which are inconsistent with normal Fermi-liquid (FL) behavior. This includes the {\em linear} dependence of the scattering rates on temperature ($T$) or energy ($\omega$) which has been suggested \cite{Varma} as characterizing a non-FL state, known as the ``marginal FL'' (MFL) state. Similar features were also observed in the iron-based superconductors (FeSCs), in which HTSC was discovered later, and in other SC materials referred to here as ``unconventional''. These anomalous features have been attributed to quantum criticality \cite{Castro, Sachdev, Zaanen}, as is happening in magnetic insulators \cite{Sachdev}. In such cases, there exists a $T=0$ quantum phase transition under the variation of a parameter $p$; for $p>p_c$ the ground state is homogeneous while for $p<p_c$ it is of an ordered magnetic state (e.g., an antiferromagnet (AF)) which breaks the symmetry between different possible spin arrangements and directions. At the quantum critical point $(p,T) = (p_c,0)$, and a $T>0$ critical regime \cite{Sachdev}, the quantum state is a complex combination of states with macroscopically different symmetries. In attempts to attribute a similar type of phase diagram to the cuprates, $p$ generally represents extra charge introduced to the parent compound through doping. 

The coupling-strength parameter $g$ is generally believed to be large in the underdoped regime (thus for small values of $p$) and small in the overdoped regime (thus large $p$). A generic behavior of $T_c$ in such unconventional superconductors, as a function of $g$,  is sketched in Fig.~\ref{fig1}(b). This demonstrates how the existence of a quantum critical regime extends the range of values of $g$ in which $T_c$ can be increased with it, allowing for HTSC to occur. Note that the boundaries of the critical regime in Fig.~\ref{fig1}(b) are broad, and no phase transitions are expected to take place there; thus the broken-symmetry regime neighboring it is expected to be of a short-range/dynamical nature, and a phase transition to a long-range broken-symmetry state would take place (if possible) towards $T=0$ \cite{Sachdev}. 

A number of theoretical approaches have been proposed in attempts to understand the coincidence between quantum criticality and HTSC, viewed in Fig.~\ref{fig1}(b)  \cite{Deutscher, Zaanen1, Zaanen2, Jarrell2, Phillips}. In a recent work by the authors \cite{AshkJohn} it was shown using a large-$U$ ansatz that even though superconductivity is suppressed within a symmetry-breaking state, the fact that the different broken-symmetry states are combined within the critical regime activates the very-strong-coupling pairing glue, allowing for HTSC to occur. This work,  referred to as the auxiliary Bose condensates (ABC) approach, is  related to a previous work by one of the authors \cite{AshkHam, Ashkenazi}, where the existence of a  combination state of striped states of different symmetries has been assumed for the hole-doped cuprates; however, no association was  made in that work to the fact that quantum criticality is the driving force behind the stabilization of the combination state \cite{Sachdev}.

\section{Formalism}
The minimal reduced Hamiltonian necessary to investigate the universal low-energy properties of the cuprates, is \cite{Ashkenazi} a one-band 2D Hubbard model on a square lattice (unit vectors $a{\hat x}$ and $a{\hat y}$), with transfer integrals up to the third-nearest neighbor (i.e. including the parameters $t$, $t^{\prime}$ and $t^{\prime\prime}$). The on-site Coulomb repulsion parameter $U$ within this band corresponds to the large-to-intermediate-$U$ regime. A dynamical cluster quantum Monte Carlo study of a model of this type \cite{Jarrell1, Jarrell3} confirmed the existence of a quantum-critical MFL regime, between the pseudogap (PG) and FL regimes. A related attempt \cite{Jarrell2} to prove that the calculated $T_c$ is raised within the quantum critical regime, was less conclusive \cite{Jarrell3}. The theory upon which Ref.~\cite{Jarrell2} is based, assumes the pairing of electron-like quasiparticles (QPs) due to their coupling to spin fluctuations. Such QPs correspond to a FL state and may be expected to describe the heavily overdoped regime -- i.e. the small-$g$ range in  Fig.~\ref{fig1}(b) -- but it is more doubtful for the lower-doping (i.e. larger-$g$) non-FL regimes.

Within the ABC approach \cite{AshkJohn, AshkHam, Ashkenazi}, this Hamiltonian is treated by the large-$U$ auxiliary-particle method \cite{Barnes} which should provide a reasonable description of the  low-energy properties of the cuprates within, at least, most of the doping range where SC occurs. (Hence it may be excluded within the range of very small values of $g$ in Fig.~\ref{fig1}(b)). Using this approach, QPs are derived on the basis of auxiliary particles which can be described \cite{AshkHam} as combinations of atomic-like configurations with the same number of electrons per site, and a constraint must be imposed \cite{Barnes} to maintain the physical situation of one configuration per site. Hence beside the cuprates, this approach is applicable to low-energy excitations in multi-band systems such as the FeSCs. We apply the existing freedom to choose configurations corresponding to the number of electrons per site in the undoped ($p = 0$) case, to be bosons -- we call these svivons. Those corresponding to one electron more, or less, per site are then fermions, called here quasi-electrons (QEs).  

In the cuprates, this choice of the statistics of the QPs corresponds to the ``slave fermion'' method -- not the ``slave boson'' method (including the Gutzwiller approximation) which characterizes most RVB models. However unlike the mean-field approach within which the above constraint was generally maintained in previous works,  the ABC approach takes into account the presence of dynamical inhomogeneities, and maintains them in a site-dependent and time-dependent manner. Consequently, a dynamical field of Lagrange multipliers is introduced which is equivalent \cite{Ashkenazi} to an additional field of bosons, called here lagrons. Within a grand-canonical scheme, the Hamiltonian includes QE--lagron and svivon--lagron coupling terms which, due to the nature of the constraint, represent the coupling of electrons to spin, orbital and charge fluctuations. Thus, works treating the constraint through a mean-field approximation disregard significant physical effects due to the coupling of the electrons to such fluctuations. 

Each of these three QPs has a distinct yet major role in the physics of the system: the svivons are mainly involved in the establishment of spin/orbital/charge inhomogeneities, due to their Bose condensation; the doping-induced fermion QEs are the major QPs behind charge dynamics; the lagrons play the role of the coupling bosons, where their coupling to QEs induces the pairing of the charge carriers, and their coupling to svivons underlies the symmetry-breaking inhomogeneities.

\section{Results}
Theoretical studies \cite{Emery1} on the basis of similar Hamiltonians, predict that the interplay between electron hopping and AF exchange in the hole-doped cuprates is likely to drive the formation of a striped structure. For doping levels $p$ where SC could occur, such structures have been observed under special doping conditions, e.g., including Nd \cite{Tran1}, or close to ``1/8 anomalies" \cite{Tran2}, and their occurrence results in a substantial decrease in $T_c$. They are  characterized by a spin density wave (SDW) of wave vector:
\begin{equation}
{\bf Q}_m = {\bf Q} + \delta {\bf q}_m,\ {\rm for} \ m = 1\ {\rm or}\  2\ {\rm or} \  3\ {\rm or} \ 4, \ \ \ \label{eq1} 
\end{equation}
where ${\bf Q} = (\pi/a)({\hat x}+{\hat y})$ is the wave vector of the AF order in the parent compounds, and $\delta {\bf q}_m = \pm \delta q {\hat x} \ {\rm or} \ \pm \delta q {\hat y}$ are modulations around ${\bf Q}$ (typically $\delta q \cong \pi/4a$). 

Similarly to an ordered state in magnetic insulators \cite{Sachdev}, such a striped state breaks the symmetry between different possible spin arrangements and directions. Within the ABC approach, this state corresponds to a Bose-condensed svivon field whose spectrum has a V-shape energy minimum $k_{_{\rm B}} T /\mathcal{O}(N)$ at the points $\pm {\bf Q}_m /2$ \cite{AshkHam, Ashkenazi}, for a possible value of $m$ in Eq.~(\ref{eq1}), and choice of ${\bf Q}/2$ (there are four inequivalent possibilities for choosing ${\bf Q}/2$ \cite{Ashkenazi}). Consequently, gaps open up on a part of the QE and electron FS, and time-reversal symmetry is removed, resulting (as was discussed above) in the suppression of SC pairing. This consequence is prevented within a combination state of different possible svivon Bose condensates determined through Eq.~(\ref{eq1}). Each condensate then stops being an eigenstate, the svivon energy minima rise above zero and become parabolic \cite{AshkHam}, and an overall time-reversal symmetry is reinstated.  The combination state is generated through svivon--lagron coupling, where the lagron spectrum has \cite{Ashkenazi} V-shape minima of energy $k_{_{\rm B}} T /\mathcal{O}(N)$ at the four ${\bf Q}_m$ points in Eq.~(\ref{eq1}).

Electron states are convoluted QE--svivon states, hence their normal Green's-function matrix $\underline{\cal G}^d$  is derived \cite{Ashkenazi} from a ``bare'' matrix $\underline{\cal G}^d_0$ due to bubble diagrams of QE and svivon Green's functions. However there exists a significant ``dressing'' process of $\underline{\cal G}^d_0$  through QE--svivon scattering processes induced by an inter-site  transfer matrix $\underline{\tilde t}$. (These three matrices are diagonal in the ${\bf k}$ representation). Multiple-scattering introduces a self-energy correction $\underline{\Sigma}^d \cong \underline{\tilde t} (\underline{1} - \underline{\cal G}^d_0 \underline{\tilde t})^{-1}$, yielding:
\begin{equation} 
\underline{\cal G}^d \cong \big( \underline{1} - \underline{\cal G}^d_0
\underline{\tilde t} \big) \big( \underline{1} - 2\underline{\cal G}^d_0 \underline{\tilde t} \big)^{-1} \underline{\cal G}^d_0. \label{eq6} 
\end{equation}
Consequently, the normal-state electron Green's functions have two types of poles: ({\it i}) QE-svivon convolution introduces a continuity of poles per ${\bf k}$ state, with maximal contribution from svivons close to their energy minima, and thus a non-FL-type feature; ({\it ii}) the multiple-scattering dressing term  $(\underline{1} - 2\underline{\cal G}^d_0 \underline{\tilde t})^{-1}$ in  Eq.~(\ref{eq6})  introduces one pole per ${\bf k}$ state, and thus a FL-type feature. The quantum state of normal-state electrons is therefore a combination of a FL state and non-FL states of different broken-symmetry striped structures, characteristic of a quantum critical regime \cite{Sachdev}. Indeed, the observed critical regime appears projected from a $T=0$ quantum phase transition which is missing in the presence of the SC regime. This hidden transition is between a low-$p$ broken-symmetry non-FL striped state, and a high-$p$ homogeneous FL state. 

Since the broken-symmetry striped states correspond to svivon condensates, their combination restores a higher symmetry {\it vis a vis} the QEs. The evaluation of the QE spectrum, and that of the electron, is based on an adiabatic treatment of the combined states. The resulting QE spectrum \cite{Ashkenazi} includes the AF wave vector ${\bf Q}$ as a reciprocal lattice vector, due to averaging over points ${\bf Q}_m$ in Eq.~(\ref{eq1}). This is reflected in the existence of equivalent main and shadow QE bands. The equivalence between these bands is removed in the resulting electron spectrum, but still both bands exist. Signatures of the ${\bf Q}_m$ wave vectors are expected to persist in cases of slow stripes dynamics. The QE spectrum \cite{Ashkenazi} consists of Brillouin-zone (BZ) areas of flat, polaron like, low-energy QEs, and an abrupt transition between them and the higher-energy QE bands in the rest of the BZ. This feature is reflected in the electron bands through the existence of kinks close the Fermi level ($E_{_{\rm F}}$) followed by ``waterfalls'' further away from $E_{_{\rm F}}$. 

\begin{figure}  
\centering
\includegraphics[width=\columnwidth]{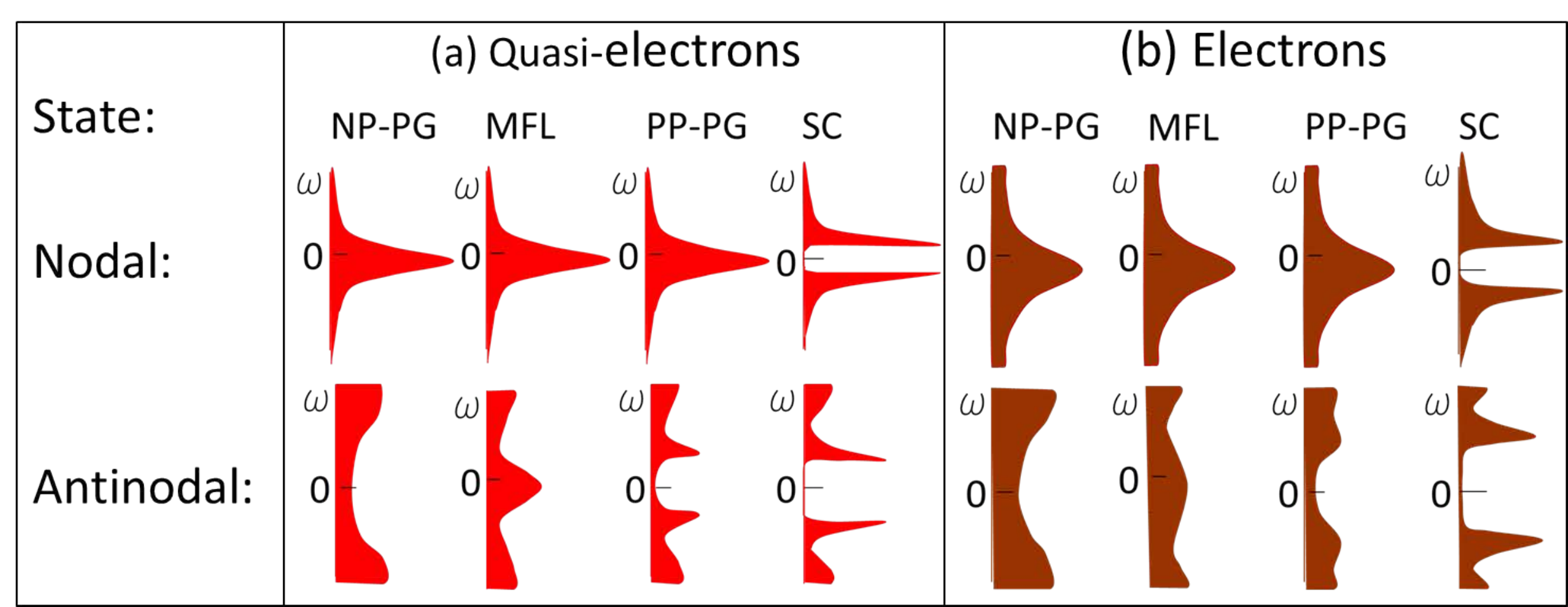}
\caption{Typical low-energy (a) QE and (b) electron spectral functions in the NP-NG, MFL,PP-PG and SC states, at nodal and antinodal points in hole-doped cuprates. The antinodal QEs are paired both in the PP-PG and the SC states, while nodal QEs are paired only in the SC state.}
\label{fig2}
\end{figure}

For a static striped state  (i.e. a SDW) of one of the ${\bf Q}_m$ wave vectors in Eq.~(\ref{eq1}), a zero-energy gap opens up at low $T$ in ${\bf k}$ points of low-energy QEs, if there are low energy QEs also close to ${\bf k} + {\bf Q}_m$ or ${\bf k} - {\bf Q}_m$. This results in a split-peak QE spectral structure around such ${\bf k}$ points, and a one-peak QE spectral structure at other low-energy ${\bf k}$ points. Within the regime where short-range stripe-like inhomogeneities exist (which corresponds to the low short-range symmetry zone  in  Fig.~\ref{fig1}(b)), their effects are averaged throughout the sample. Consequently, the traces of the above peak structures are manifested as two wide QE spectral peaks (``humps'') at a positive and a negative energy, around the antinodal BZ areas, and a single peak around the nodal areas. Such a scenario is expected within the range referred to here as the no-pairing pseudogap (NP-PG) state of the hole-doped cuprates. The corresponding QE spectral features are sketched in  Fig.~\ref{fig2}(a), and their reflection in the low-energy electron spectrum  in  Fig.~\ref{fig2}(b).
 
In contrast to what happens in the NP-PG regime, for a combination state of striped states of the different ${\bf Q}_m$ wave vectors occurring within the critical regime (thus in the MFL state), a coherent combination around the antinodal BZ areas results \cite{Ashkenazi} in a three-peak structure at  low-QE-energy ${\bf k}$ points. On the other hand, the single-peak scenario prevails around the nodal BZ areas. The resulting QE spectral features, within the MFL state, are sketched  in  Fig.~\ref{fig2}(a), and their reflection in the low-energy electron spectrum  in  Fig.~\ref{fig2}(b). Since each peak in the QE spectral functions corresponds to a pole in their Green's functions, the QE BZ is divided within the critical regime into two analytically distinct $T$-dependent areas, with one pole per ${\bf k}$ point in one and three poles in the other. 

The derived \cite{Ashkenazi} electron scattering rates  $\Gamma^d({\bf k},\omega,T)$ are dominated for  small and intermediate values of $\omega$ and $T$ by terms of the form: 
\begin{equation}
\Gamma^d({\bf k},\omega,T) \simeq \Gamma^d_0({\bf k}) - \Gamma^d_1({\bf k}) |\omega| b_{_T}(-|\omega|),  \label{eq2} 
\end{equation}
where $b_{_T}(\omega)\equiv 1/[\exp{(\omega/k_{_{\rm B}}T)}-1]$ is the Bose distribution function which emerges from the QE--svivon convolution expression. Eq.~(\ref{eq2}) yields: 
\begin{equation}
\Gamma^d({\bf k},\omega,T) \simeq \begin{cases} \Gamma^d_0({\bf k}) + \Gamma^d_1({\bf k}) k_{_{\rm B}}T,& \text{for $|\omega| \ll k_{_{\rm B}}T$,}\\ \Gamma^d_0({\bf k}) + \Gamma^d_1({\bf k})|\omega|,& \text{for $|\omega| \gg k_{_{\rm B}}T$,} \end{cases}
\label{eq3}
\end{equation}
in agreement with MFL phenomenology \cite{Varma}. This further demonstrates that our  description is realistic for a quantum critical state.

The linear variation of $\Gamma^d({\bf k},\omega,T)$ with $\omega$ and $T$ is derived from the QE--svivon convolution expression, and the scattering processes determining it are mostly due to QE--lagron and svivon--lagron coupling. The coupling of QEs to lagrons also induces their pairing, under which the electrons become paired as well. Since, as mentioned above, the BZ of the low-energy QEs consists (within the critical regime) of two analytically distinct parts, pairing can occur below different temperatures, $T_p$ and $T_c$ ($T_p \ge T_c$) in the antinodal and nodal areas respectively. The SC state corresponds to the $T < T_c$ regime. The $T_c < T < T_p$ regime is referred to as the partial-pairing pseudogap (PP-PG) state. Note that the range which is commonly regarded as the PG regime includes here this PP-PG state and also the NP-PG state discussed above. The QE spectral features in the PP-PG and SC states are sketched  in  Fig.~\ref{fig2}(a), and their reflection in the low-energy electron spectrum  in  Fig.~\ref{fig2}(b).

\section{Discussion}
The partial pairing around the antinodal BZ areas in the PP-PG state does not result in supercurrent carrying Cooper pairs (which requires scattering without breaking between different $({\bf k}, -{\bf k})$ states). Within the PP-PG regime QE pairs are broken when they are scattered between the antinodal and the nodal BZ areas, but they are still expected to introduce a signal of diamagnetism. Indeed, Nernst-effect \cite{Ong} and spectral-weight \cite{Kondo} measurements confirm that the commonly considered PG regime includes a partial range where some pairing occurs (see discussion in Ref.~\cite{Marel7}). In addition to the above, transport \cite{Alloul} and terahertz spectroscopy \cite{Bozovic} results confirm the existence of a range of SC fluctuations above $T_c$ which does not coincide with either of the above PG states. $T_c$ could be raised, extending the SC regime on account of this fluctuations range, by reducing the degree of disorder, misfit strains, {\it etc.} \cite{Pavuna, Fratini, Poccia, Johnson}.

\begin{figure}  
\centering
\includegraphics[width=\columnwidth]{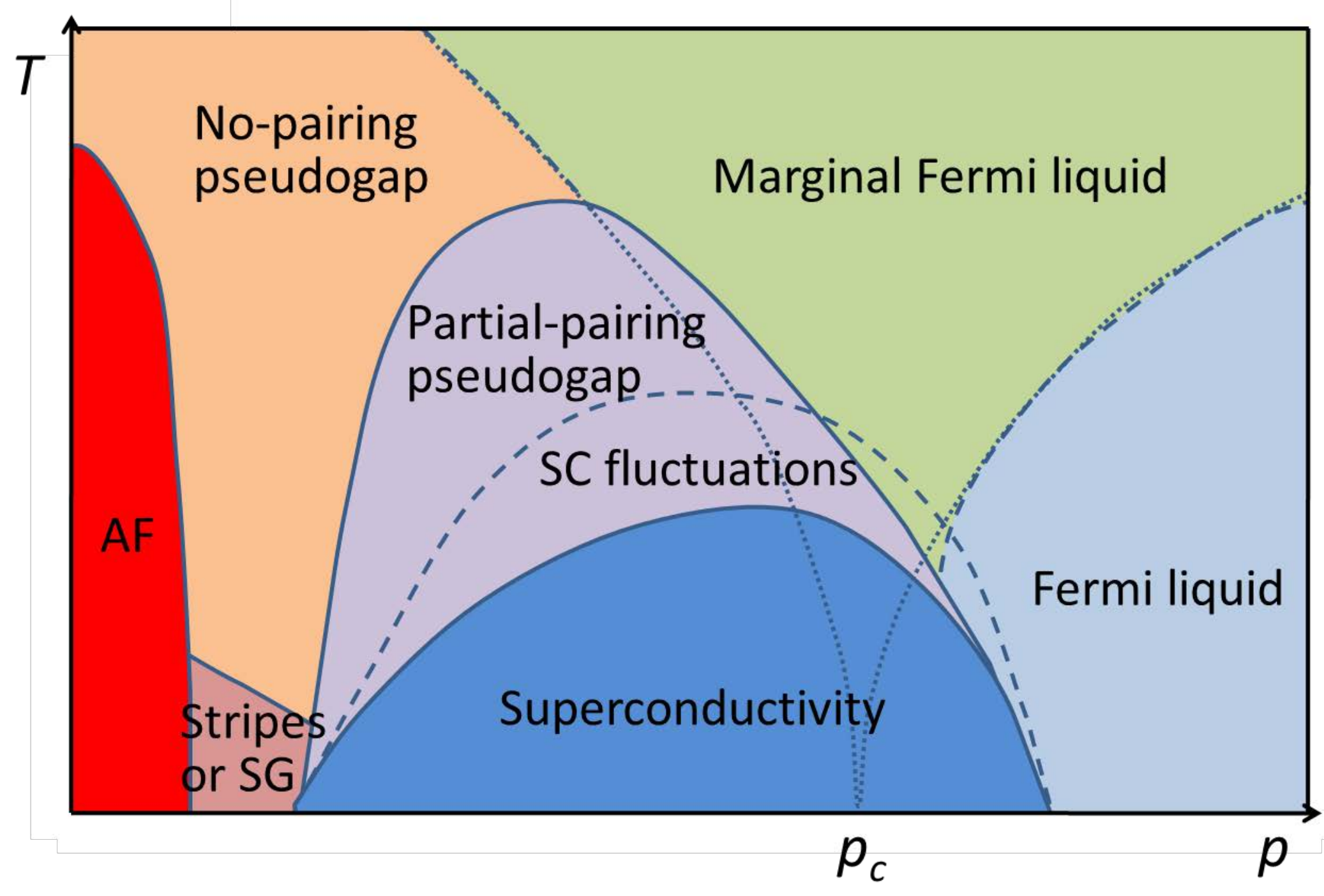}
\caption{Proposed generic phase diagram for the hole-doped cuprates, under changing temperature and doping level $p$. Solid and dashed lines represent phase and gradual transitions, respectively. Dotted lines represent gradual transitions in the case that pairing does not occur, or is suppressed. The acronym SG corresponds to a spin glass.} 
\label{fig3}
\end{figure}

The resulting generic phase diagram of hole-doped cuprates is sketched in Fig.~\ref{fig3}. In the case that pairing does not occur, or is suppressed, one gets (similarly to magnetic insulators \cite{Sachdev}) a quantum critical point at $(p,T) = (p_c,0)$, and a critical regime between the dotted lines in Fig.~\ref{fig3}. However, since the establishment of the PP-PG and SC states requires the combination of symmetry-breaking states, existing within the quantum critical regime, the pairing-energy gain in these states results in the {\em extension} of the critical regime (in the sense of being characterized by a combination state) into the PP-PG and SC ranges in Fig.~\ref{fig3}. Due to the pairing in these states, the transitions between them and their neighboring ranges are phase transitions. For $p<p_c$, pairing {\it within} the PP-PG and SC states can be suppressed under specific doping conditions \cite{Tran1, Tran2}, restoring the broken-symmetry non-FL striped state.

The existence of the PP-PG state introduces a modification of this phase diagram in comparison to the generic behavior of unconventional superconductors, presented in Fig.~\ref{fig1}(b). The MFL regime is what remains of the unextended quantum critical regime, and consistently with Fig.~\ref{fig1}(b), the transitions between it and the NP-PG (low short-range symmetry) and FL (high symmetry) regimes are gradual. A manifestation of the gradual transition between the MFL and the FL regimes is the observed crossover between the linear behavior of the scattering rates (see Eq.~(\ref{eq3})) and a quadratic behavior characteristic of a FL. Typically for a gradual transition, remnants of MFL behavior persist within the normal state in the overdoped regime \cite{Kokalj}. As can be seen in  Fig.~\ref{fig3}, there exists a range of the phase diagram where a direct FL--SC transition occurs. This heavily overdoped regime may constitute a crossover between the ranges of applicability of the ABC approach (under which this transition traverses the boundary of the extended critical regime) and of a model of pairing of electron-like QPs in a FL, due to their coupling to spin fluctuations \cite{Jarrell2, Scalapino1,  Maiti}. Time-of-flight neutron spectroscopy measurements \cite{Wakimoto} of the magnetic excitations, throughout the entire doping regime, confirm a gradual disappearance of signals related to stripe fluctuations, when $p$ is increased from the deeply underdoped to the deeply overdoped regime as predicted here.

 Note that the process of coupling of electrons to spin/orbital/charge fluctuations, represented here through svivon--lagron and QE--lagron coupling, results in the striped states and causes pairing when these states are combined due to quantum criticality. Hence the lagrons are playing the role of a {\em critical pairing glue} which is activated in the critical regime (extending into the SC and PP-PG ranges in Fig.~\ref{fig3}). It is deactivated in states of long- or short-range striped/AF structures occurring within the NP-PG, stripes/SG and AF regimes in Fig.~\ref{fig3} (note that for very low $p$  such inhomogeneous structures are {\em not} characterized by the wave vectors in Eq.~(\ref{eq1}) \cite{Yamada}), or under specific doping conditions \cite{Tran1, Tran2}. Deep within the FL regime, the assumptions of the ABC approach (introducing QEs, svivons and lagrons as QPs) stop being valid, and thus this critical pairing glue fades away.

Within the MFL regime in Fig.~\ref{fig3}, stripe dynamics are fast and the short lifetimes of the combined striped structures yield features which are too broad to be observed experimentally. This changes when at least a partial gap opens up in the NP-PG, PP-PG and SC regimes. Indeed, the existence of fluctuating stripes has been observed in these regimes using STM \cite{Yazdani}, and the resulting broken rotational symmetry through the anisotropic Nernst effect \cite{Daou}. The behavior of the fluctuating stripes has been analyzed by analogy to nematicity in liquid crystals \cite{Kivelson, Lawler, Carlson}. The existence of such fluctuating SDW structures implies that experiments measuring timescales shorter than the stripe dynamics will detect apparent violations of time-reversal symmetry, while longer timescale experiments will not. Indeed,  neutron scattering measurements \cite{Bourges1} seem to indicate such violations within the PP-PG regime, while Zeeman-perturbed NQR \cite{Keller1} and zero-field $\mu$SR \cite{Huang}  results  do not indicate it. For {\em fluctuating} stripes within a combination state, an over-all time-reversal symmetry is maintained, and thus pairing is {\em not} suppressed.

Similar predictions follow for other unconventional SC systems, including a quantum-critical regime. The key feature is that the pairing bosons -- which could be of a similar, or a different, nature to the ones in the hole-doped cuprates -- introduce a symmetry-breaking instability, and these broken-symmetry states are combined in the critical regime, restoring symmetry for the paired QPs. Similar QPs and pairing bosons to those in the cuprates, are expected for the FeSCs -- however they will be more evolved due to their multi-band nature. Our theory predicts that the quantum phase transition in the FeSCs is between a non-FL magnetic state and a FL state, and that the broken-symmetry states are the perpendicular-direction SDWs \cite{Lynn1} which exist, or almost exist, in the magnetic state. The expected short-range order regime in the FeSCs is of nematicity characterized by both spin and orbital features. In the electron-doped cuprates, this regime is of short-range AF order. 

\section{Conclusions}
Regarding searches for new HTSC systems, the message of the present work is that a good place to look for them is in multi-element compounds and alloys, at the vicinity of phase transitions, or transitions in the nature of the electrons around $E_{_{\rm F}}$ between  large-$U$ and smaller-$U$ characteristics. This includes Mott transitions, and transitions in multi-band compounds where large-$U$ electrons could reside at the close vicinity of $E_{_{\rm F}}$. 

The fact that $T_c$ is higher in the hole-doped cuprates than in the electron-doped cuprates and in the FeSCs, may be connected with the fact that the degeneracy of the broken-symmetry states in the hole-doped cuprates is higher than in the two other systems. Hence they are striped states specified by four inequivalent wave vectors ${\bf Q}_m$ (see Eq.~(\ref{eq1})), rather than by one or two wave vectors, as in the other systems. This results \cite{Ashkenazi} in a critical-regime combination of the striped states with a wider range of low-energy lagrons -- in other words, there is a larger phase space for {\em soft} pairing bosons representing spin/orbital/charge fluctuations. A reasonable conclusion concerning the search for HTSC systems of a higher $T_c$ than in the hole-doped cuprate, is that the route to them is through more evolved large-$U$ systems with a {\em higher degeneracy} of symmetry-breaking states. These states, which are combined in the critical regime, should be characterized by a larger number of inequivalent wave vectors than the striped states in the hole-doped cuprates.


\begin{thebibliography}{1}
\bibitem{Testardi}L.R.~Testardi, {\it Rev.~Mod.~Phys.} {\bf 47}, 637 (1975).
\bibitem{Kresin}V.Z.~Kresin, and S.A.~Wolf, {\it Rev.~Mod.~Phys.} {\bf 81}, 481 (2009). 
\bibitem{Varma}C.M.~Varma, P.B.~Littlewood, S.~Schmitt-Rink, E.~Abrahams and A.E.~Ruckenstein, {\it Phys.~Rev.~Lett.} {\bf 63}, 1996 (1989).
\bibitem{Castro}C.~Castellani, C.~Di Castro, and M.~Grilli, {\it Physica C} {\bf 282--287}, 260 (1997).
\bibitem{Sachdev}S. Sachdev and B. Keimer, {\it Physics Today} {\bf 64}, issue No.~2, 29 (2011), and references therein.
\bibitem{Zaanen}J. Zaanen, in {\it 100 years of superconductivity}, ed. H.~Rogalla and P.~H.~Kes (Taylor \& Francis, p. 92, 2011).
\bibitem{Deutscher}G.~Deutscher and Y.~Dagan, {\it J.~Supercond.~Nov.~Mag.} {\bf 13}, 699 (2000).
\bibitem{Zaanen1}J.-H.~She, and J.~Zaanen, {\it Phys.~Rev.~B} {\bf 80}, 184518 (2009).
\bibitem{Zaanen2}J.-H.~She, B.J.~Overbosch, Y.-W.~Sun, Y.~Liu, K.~Schalm, J.A.~Mydosh, and J.~Zaanen, {\it Phys.~Rev.~B} {\bf 84}, 144527 (2011).
\bibitem{Jarrell2}S.-X.~Yang, H.~Fotso, S.-Q.~Su, D.~Galanakis, E.~Khatami, J.-H.~She, J.~Moreno, J.~Zaanen, and M.~Jarrell, {\it Phys.~Rev.~Lett.} {\bf 106}, 047004 (2011).
\bibitem{Phillips}M.~Edalati, R.~G.~Leigh, K.~W.~Lo, and P.~W.~Phillips, {\it Phys.~Rev.~D} {\bf 83}, 045012 (2011). 
\bibitem{AshkJohn}J.~Ashkenazi, and N.F.~Johnson, {\it Europhys.~Lett.}  {\bf 98}, 47011 (2012).
\bibitem{AshkHam}J.~Ashkenazi, {\it J.~Supercond.~Nov.~Magn.} {\bf 22},
3 (2009).
\bibitem{Ashkenazi}J.~Ashkenazi, {\it J.~Supercond.~Nov.~Mag.} {\bf 24}, 1281 (2011). 
\bibitem{Jarrell1}N.S.~Vidhyadhiraja, A.~Macridin, C.~Sen, M.~Jarrell, and M.~Ma, {\it Phys.~Rev.~Lett.} {\bf 102}, 206407 (2009).
\bibitem{Jarrell3}K.-S.~Chen, S.~Pathak, S.-X.~Yang, S.-Q.~Su, D.~Galanakis, K.~Mikelsons, M.~Jarrell, and J.~Moreno, {\it Phys.~Rev.~B} {\bf 84}, 245107 (2011).
\bibitem{Barnes}S.E.~Barnes, {\it Adv.~Phys.} {\bf 30}, 801 (1981). 
\bibitem{Emery1}V.J.~Emery, and S.A.~Kivelson, {\it Physica C} {\bf
209}, 597 (1993).
\bibitem{Tran1}J.M.~Tranquada, J.D.~Axe, N.~Ichikawa, Y.~Nakamura,
S.~Uchida, and B.~Nachumi, {\it Phys.~Rev.~B} {\bf 54}, 7489 (1996).
\bibitem{Tran2}M.~Fujita, H.~Goka, K.~Yamada, J.M.~Tranquada, and L.P.~Regnault, {\it Phys.~Rev.~B} {\bf 70}, 104517 (2004).
\bibitem{Ong}L.~Li, Y.~Wang, S.~Komiya, S.~Ono, Y.~Ando, G.~D.~Gu, and N.~P.~Ong,  {\it Phys.~Rev.~B} {\bf 81}, 054510 (2010).
\bibitem{Kondo}T. Kondo, Y.~Hamaya, A.~D.~Palczewski, T.~Takeuchi, J.~S.~Wen, Z.~J.~Xu, G.~Gu, J.~Schmalian, and A.~Kaminski,  {\it Nature Physics} {\bf 7}, 21 (2011).
\bibitem{Marel7}Dirk van der Mareli,  {\it Nature Physics} {\bf 7}, 10 (2011).
\bibitem{Alloul}H.~Alloul, F.~Rullier-Albenque, B.~Vignolle, D.~Colson, and A.~Forget, {\it Europhys.~Lett.} {\bf 91}, 37005 (2010). 
\bibitem{Bozovic}L.~S. Bilbro, R.~V.~Aguilar, G.~Logvenov, O.~Pelleg, I.~Bozovic, and N.~P.~Armitage,  {\it Nature Physics} {\bf 7}, 298 (2011).
\bibitem{Pavuna}D.~Pavuna, D.~Ariosa, C.~Cancellieri, D.~Cloetta, and M.~Abrecht, {\it Journal of Physics} {\bf 108}, 012040 (2008). 
\bibitem{Fratini}M.~Fratini, N.~Poccia, A.~Ricci, G.~Campi, M.~Burghammer, G.~Aeppli, and A.~Bianconi, {\it  Nature} {\bf 466}, 841 (2010). 
\bibitem{Poccia}N.~Poccia, A.~Ricci and A.~Bianconi, {\it J.~Supercond.~Nov.~Mag.} {\bf 24}, 1195 (2011).
\bibitem{Johnson}N.F.~Johnson, J.~Ashkenazi, Z.~Zhao, and L.~Quiroga, {\it AIP Advances} {\bf 1}, 012114 (2011).
\bibitem{Kokalj}J.~Kokalj and R.H.~McKenzie, {\it Phys.~Rev.~Lett.} {\bf 107}, 147001 (2011).
\bibitem{Scalapino1}N.~E.~Bickers, D.~J.~Scalapino and S.~R.~White, {\it Phys.~Rev.~Lett.} {\bf 62}, 961 (1989). 
\bibitem{Maiti}S.~ Maiti, M.~M.~Korshunov, T.~A.~Maier, P.~J.~Hirschfeld, and A.~V.~Chubukov,  {\it Phys.~Rev.~B} {\bf 84}, 224505 (2011).
\bibitem{Wakimoto}S.~Wakimoto, K.~Yamada, J.~M.~Tranquada, C.~D.~Frost, R.~J.~Birgeneau, and H.~Zhang, {\it Phys.~Rev.~Lett.} {\bf 98}, 247003 (2007).
\bibitem{Yamada} K.~Yamada, C.~H.~Lee, K.~Kurahashi, J.~Wada, S.~Wakimoto, S.~Ueki, H.~Kimura, Y.~Endoh, S.~Hosoya, G.~Shirane, R.~J.~Birgeneau, M.~Greven, M.~A.~Kastner, and Y.~J.~Kim, {\it
Phys.~Rev.~B} {\bf 57}, 6165 (1998). 
\bibitem{Yazdani}C.V.~Parker, P.~Aynajian, E.H.~da Silva Neto, A.~Pushp, S.~Ono, J.~Wen, Z.~Xu, G.~Gu, and A.~Yazdani, {\it Nature} {\bf 468}, 677 (2010). 
\bibitem{Daou}R.~Daou, J.~Chang, D.~LeBoeuf, O.~Cyr--Choini\`ere, F.~Lalibert\'e, N.~Doiron--Leyraud, B.~J.~Ramshaw, R.~Liang, D.A.~Bonn, W.N.~Hardy, and L.~Taillefer, {\it Nature} {\bf 463}, 519 (2010). 
\bibitem{Kivelson}S.~A.~Kivelson, E.~Fradkin, and V.~J.~Emery, {\it Nature} {\bf 393}, 560 (1998). 
\bibitem{Lawler}M.~J.~Lawler, K.~Fujita, J.~Lee, A.~R.~Schmidt, Y.~Kohsaka, C.-K.~Kim, H.~Eisaki, S.~Uchida, J.~C.~Davis, J.~P.~Sethna, and E.-A.~Kim, {\it Nature} {\bf 466}, 347 (2010). 
\bibitem{Carlson}E.~W.~Carlson and K.~A.~Dahmen, {\it Nature Commun.} {\bf 2}, 379 (2011).
\bibitem{Bourges1}Y.~Li, V.~Bal\'edent, G.~Yu, N.~Barisi\'c, K.~Hradil, R.A.~Mole, Y.~Sidis, P.~Steffens, X.~Zhao, P.~Bourges, and M.~Greven, {\it Nature} {\bf 468}, 283 (2010).
\bibitem{Keller1}S.~Str\"assle, B.~Graneli, M.~Mali, J.~Roos1, and H.~Keller, {\it Phys.~Rev.~Lett.} {\bf 106}, 097003 (2011). 
\bibitem{Huang}W.~Huang, V.~Pacradouni, M.~P.~Kennett, S.~Komiya, and J.~E.~Sonier,  {\it Phys.~Rev.~B} {\bf 85}, 104527 (2012).
\bibitem{Lynn1} C.~de la Cruz, Q.~Huang, J.W.~Lynn, J.~Li, W.~Ratcliff II, J.L.~Zarestky, H.A.~Mook, G.F.~Chen, J.L.~Luo, N.L.~Wang, and P. Dai, {\it Nature} {\bf 453}, 899 (2008). 

\end{thebibliography}
\end{document}